\documentclass[aip,rsi,reprint]{revtex4-1}

\pdfoutput=1
\usepackage[T1]{fontenc}
\usepackage{times}
\usepackage{dcolumn}
\usepackage{natmove}
\usepackage{graphicx}
\usepackage{natbib} 
\usepackage{amsmath}
\usepackage{amssymb}
\usepackage{gensymb}
\usepackage{textcomp}
\usepackage{hyperref}
\usepackage[usenames,dvipsnames]{color}
\usepackage{longtable}
\usepackage{array}
\newcolumntype{L}[1]{>{\raggedright\let\newline\\\arraybackslash\hspace{0pt}}m{#1}}
\newcolumntype{C}[1]{>{\centering\let\newline\\\arraybackslash\hspace{0pt}}m{#1}}
\newcolumntype{R}[1]{>{\raggedleft\let\newline\\\arraybackslash\hspace{0pt}}m{#1}}

\begin{document}


\title{Quantum dots with split enhancement gate tunnel barrier control}

\author{S. Rochette}
\thanks{S. Rochette and M. Rudolph contributed equally to this work.}
\affiliation{Institut Quantique, Universit\'e de Sherbrooke, Sherbrooke J1K 2R1, Canada}
\affiliation{D\'epartement de physique, Universit\'e de Sherbrooke, Sherbrooke J1K 2R1, Canada}

\author{M. Rudolph}
\thanks{S. Rochette and M. Rudolph contributed equally to this work.}
\affiliation{Sandia National Laboratories, Albuquerque, New-Mexico 87185, USA}

\author{A.-M. Roy}
\affiliation{D\'epartement de physique, Universit\'e de Sherbrooke, Sherbrooke J1K 2R1, Canada}

\author{M. J. Curry}
\affiliation{Department of Physics and Astronomy, University of New-Mexico, Albuquerque, New-Mexico 87131, USA}

\author{G. A. Ten Eyck}
\affiliation{Sandia National Laboratories, Albuquerque, New-Mexico 87185, USA}

\author{R. P. Manginell}
\affiliation{Sandia National Laboratories, Albuquerque, New-Mexico 87185, USA}

\author{J. R. Wendt}
\affiliation{Sandia National Laboratories, Albuquerque, New-Mexico 87185, USA}

\author{T. Pluym}
\affiliation{Sandia National Laboratories, Albuquerque, New-Mexico 87185, USA}

\author{S. M. Carr}
\affiliation{Sandia National Laboratories, Albuquerque, New-Mexico 87185, USA}

\author{D. R. Ward}
\affiliation{Sandia National Laboratories, Albuquerque, New-Mexico 87185, USA}

\author{M. P. Lilly}
\affiliation{Sandia National Laboratories, Albuquerque, New-Mexico 87185, USA}
\affiliation{Center for Integrated Nanotechnology, Sandia National Laboratories, Albuquerque, New-Mexico 87185, USA}

\author{M. S. Carroll}
\affiliation{Sandia National Laboratories, Albuquerque, New-Mexico 87185, USA}

\author{M. Pioro-Ladri\`ere}
\affiliation{Institut Quantique, Universit\'e de Sherbrooke, Sherbrooke J1K 2R1, Canada}
\affiliation{D\'epartement de physique, Universit\'e de Sherbrooke, Sherbrooke J1K 2R1, Canada}
\affiliation{Quantum Information Science Program, Canadian Institute for Advanced Research, Toronto M5G 1Z8, Canada}


\begin{abstract}
We introduce a silicon metal-oxide-semiconductor quantum dot architecture based on a single polysilicon gate stack. The elementary structure consists of two enhancement gates separated spatially by a gap, one gate forming a reservoir and the other a quantum dot. We demonstrate, in three devices based on two different versions of this elementary structure, that a wide range of tunnel rates is attainable while maintaining single-electron occupation. A characteristic change in slope of the charge transitions as a function of the reservoir gate voltage, attributed to screening from charges in the reservoir, is observed in all devices, and is expected to play a role in the sizable tuning orthogonality of the split enhancement gate structure. The all-silicon process is expected to minimize strain gradients from electrode thermal mismatch, while the single gate layer should avoid issues related to overlayers (e.g., additional dielectric charge noise) and help improve yield. Finally, reservoir gate control of the tunnel barrier has implications for initialization, manipulation and readout schemes in multi-quantum dot architectures. 

\end{abstract}

\maketitle


Silicon (Si) quantum dots (QDs) are strong contenders for the realization of spin qubits \cite{Loss1998, Zwanenburg2013}. Silicon germanium heterostructure (Si/SiGe) platforms with integrated micromagnets \cite{Pioro-Ladriere2007} have produced the highest performance qubits \cite{Wu2014,Takeda2016,Kawakami2016}, with fidelities over 99.9\%\cite{Yoneda2018}, while metal-oxide-semiconductor (MOS) platforms have also achieved fault tolerant fidelities \cite{Veldhorst2014}.

Most of the high performance systems mentioned above are enhancement mode devices comprising at least two layers of control gates. The overlapping gates ensure strong confinement and the highest electrostatic control over regions surrounding the QDs. These current multi-stack devices have therefore achieved excellent tunability, thanks in part to an independant control of reservoirs, dots and tunnel barriers through respectively dedicated gates. On the other hand, single-layer enhancement mode devices are being explored for ease of fabrication and potentially higher yield, in both Si/SiGe and MOS systems \cite{Lu2016, Studenikin2018, Jock2018, HarveyCollard2018, HarveyCollard2017}. In particular, all-silicon MOS single-layer devices are expected to avoid thermal mismatch and additional dielectric charge noise from overlayers \cite{Thorbeck2015,Zimmerman2014}. These single-layer devices generally use a single gate to form a source-dot-drain channel, relying on constrictions and lateral depletion gates to shape the confinement potential \cite{Lu2016, Tracy2013}. Reservoir filling, dot charge occupation and tunnel rates are therefore controlled differently than in multi-gate stack architectures. Various architectures and methods of tunnel barrier control impact tunability differently, and understanding those differences will influence choices of multi-QDs initialization, manipulation and readout schemes, including automatic tuning procedures \cite{Baart2016,vanDiepen2018}, as well as reproducibility, versatility and scalability of devices \cite{Frees2018}.

Here we explore a single gate stack structure featuring a split gate for dot and reservoir formation. The tunnel barrier is simply formed by the gap between the dot and reservoir gates. We investigate, in all-silicon MOS devices based on this elementary structure, how tunnel barrier control can be achieved by modulation of the reservoir gate voltage. The operation principle is studied in two variations of the layout, emphasizing some intrinsic effects brought by the use of a reservoir gate for tunnel control, in contrast with the more frequent method of control using a dedicated barrier gate directly on top of the barrier. We also define a control orthogonality metric with significance for tunability and versatility of quantum dot devices and use it to compare a split gate QD device to a multi-stack device from the literature. Finally, we conclude by examining single-electron regime characteristics and valley splitting tuning in the split gate devices.

\begin{figure}
 \includegraphics*[width = 0.8\columnwidth]{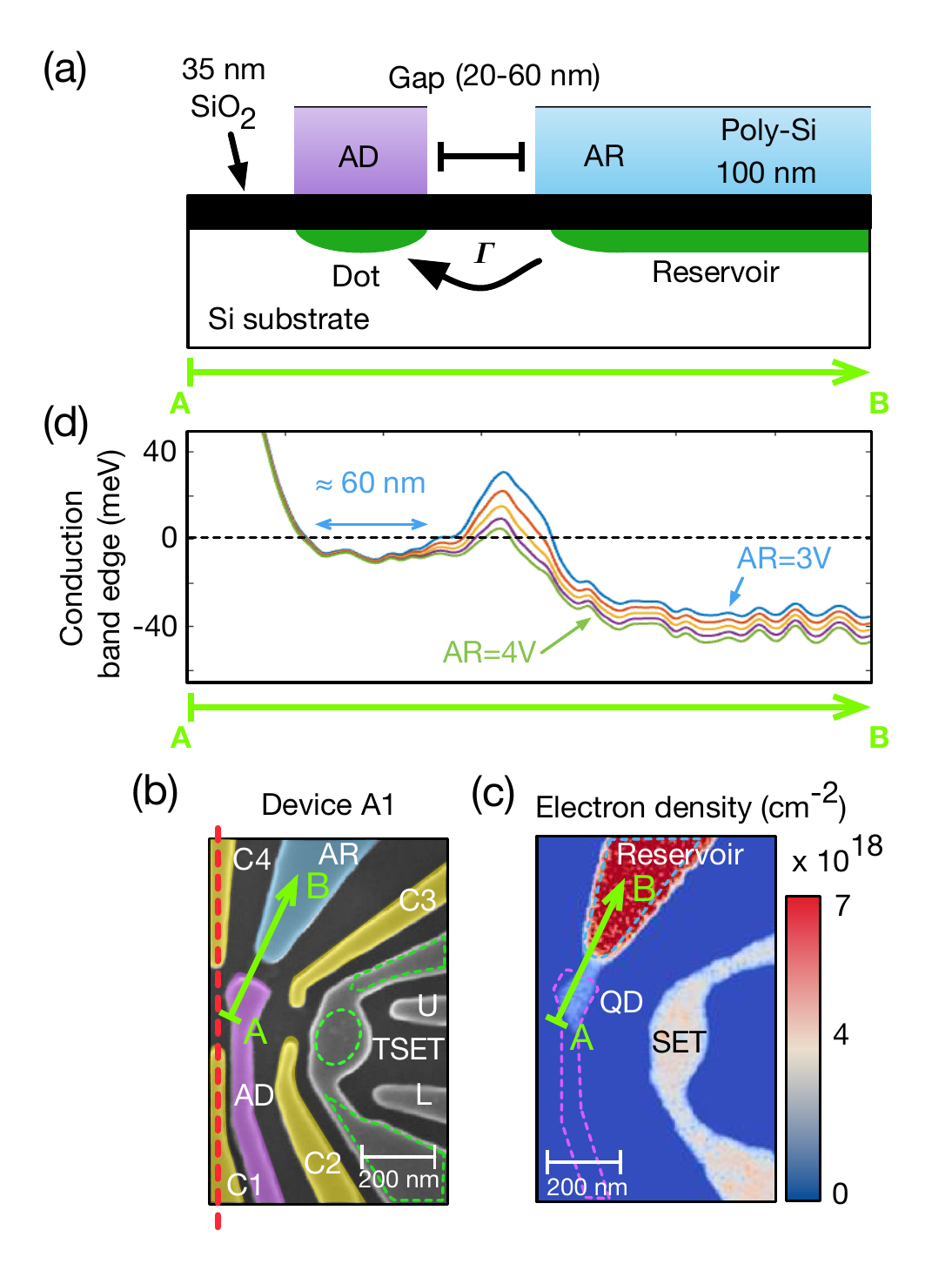}
 \caption{(a) Schematic transverse cut of the split enhancement gate tunnel barrier devices. AR is the reservoir enhancement gate, and AD is the dot enhancement gate.  (b) False-color scanning electron micrograph (SEM) of single-lead split-enhancement gate device A1. C1, C2, C3, and C4 are confinement gates. The gate TSET forms the SET channel, and U and L help define its source and drain barriers. A mirror structure, on the left side of the dotted red line, not shown for clarity, includes gates AD$'$, AR$'$, C2$'$, C3$'$, TSET$'$, U$'$, and L$'$.(c) Simulated electron density, representing approximately 20 electrons in the dot. (d) Simulated conduction band edge profile (smoothed traces) along the green arrow A-B from (a), (b), (c) and (d), for $\rm{V_{AR}}$ varying from 3 to 4 V with 0.25 V increments, with other parameters kept constant. Small amplitude modulations in the potential result from the mesh resolution used for those particular simulations and the associated sub-optimal interpolation routine.
 \label{Fig1}}
\end{figure}

The elementary single-gate stack structure we explore consists of a quantum dot enhancement gate, AD, and a reservoir enhancement gate, AR, separated by a gap, as shown in Fig.~\ref{Fig1}(a). We refer to this base unit of design as the \textit{split enhancement gate} structure. Devices are fabricated using the Sandia National Laboratories MOS quantum dot process \cite{Tracy2009,Singh2016}, which is described in detail in the Supplementary Material. The gate stack consists of a 10 000 $\Omega$-cm n-type silicon float zone substrate, a 35 nm $\rm{SiO_2}$ gate oxide and a degenerately As-doped 100 nm thick polysilicon gate (shown in Fig.~\ref{Fig1}(a)). The polysilicon nanostructure is defined by a single electron-beam lithography and dry etching step. The gate oxide properties have been characterized in Hall bars fabricated on the same starting gate stack as the nanostructures. Peak mobility, percolation density \cite{Tracy2009,Borselli2011b}, scattering charge density \cite{DasSarma2013,Tracy2009}, interface roughness and interface correlation length \cite{Mazzoni1999} were extracted for the wafers used for each of the devices and are described in the Supplementary Material.

In this study,  we look at two different layouts of split enhancement gate devices. We examine a single-lead layout (devices A1 and A2), where a single reservoir is connected to a dot, and a double-lead layout (device B), where the dot is connected in series to reservoirs to enable transport measurements, in addition to charge sensing. Devices A1 and A2 present the same layout, with only differences in scale and spacing (see Table I in the Supplementary Material). For all devices, measurements are performed using a proximal SET as a charge sensor with standard lock-in or RF reflectometry techniques\cite{Muller2012}. Details on the measurements and a list of all voltages employed are given in the Supplementary Material.  

To illustrate the split enhancement gate tunnel barrier structure and its operation, we have performed Thomas-Fermi numerical simulations \cite{Gao2013} of device A1, as shown in Fig.~\ref{Fig1}(b), using the corresponding MOS structure and operating gate voltages as input parameters. Figure~\ref{Fig1}(c) shows the simulated electron density at the Si/$\rm{SiO_2}$  interface when the device is experimentally set in a $\sim$20 electrons regime. As expected, a reservoir is formed under gate AR, and a quantum dot under the tip of gate AD, separated by the tunnel barrier region. Some form of tunnel barrier control using the reservoir gate voltage, $\rm{V_{AR}}$, is suggested by variations of the potential along the dot-reservoir axis (Fig.~\ref{Fig1}(d)). Indeed, as a function of $\rm{V_{AR}}$, the tunnel barrier potential height and width are modified, while the QD conduction band edge stays fairly constant relative to the Fermi level of the reservoir, indicating some form of \textit{tuning orthogonality} between charge occupation of the QD and tunnel rate to the reservoir (similar quantities are evoqued in Ref. \cite{Frees2018}). A sufficient tuning orthogonality would allow simultaneously for a wide range of tunnel rates $\Gamma$ and the ability to regularly tune these devices to the single electron regime. We therefore investigate this characteristic for a QD based on a split enhancement gate structure employing the reservoir gate as a knob \cite{Shirkhorshidian2017}.

 \begin{figure}
 \includegraphics*[width = 1.0\columnwidth]{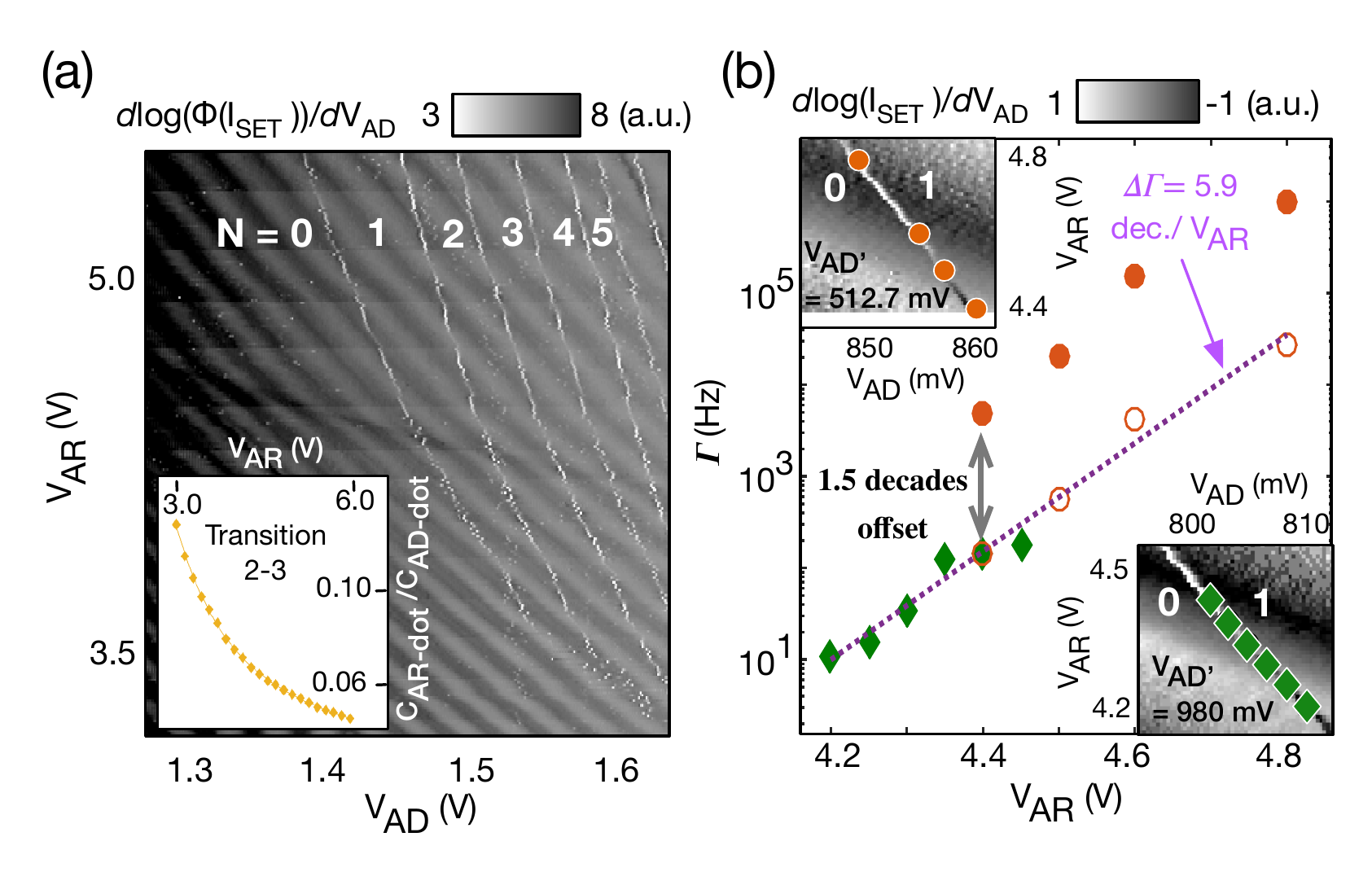}
 \caption{ (a) Stability diagram of AD vs AR in the few-electron regime for the single-lead device A2. The data was processed through a $\rm{5^{th}}$ order Butterworth digital filter and a Hilbert transform to extract the phase $\phi \rm{(SET)}$ of the signal and minimize the appearance of the background's SET's Coulomb oscillations (darker, more horizontal features). Charge occupation $N$ in the dot is indicated for each region between the transitions (thin white and more vertical features). Bottom left inset: capacitance ratio $\rm{C_{AR\text{-}dot}/C_{AD\text{-}dot}}$ as a function of $\rm{V_{AR}}$ extracted from the $N=2\rightarrow3$ charge transition's slope. (b) Reservoir-dot tunnel rate as a function of $\rm{V_{AR}}$ for the $N=0\rightarrow1$ transition in device A1. The green (diamonds) data points are obtained via full counting statistics of single-shot traces \cite{Gustavsson2009} while the orange (circles) data points are extracted from pulse spectroscopy \cite{Elzerman2004}.  Hollow orange circles are the orange filled circle data points translated by $\approx 1.5$ decades. The dotted line is an exponential fit to green and hollow orange data points, yielding a slope $\Delta\Gamma$. Top left inset: zoom on the region of the stability diagram corresponding to the orange data points, with the left dot accumulation gate AD$^{\prime}$ at 512.7 mV. Bottom right inset: Zoom on the region of the stability diagram corresponding to the green data points, with $\rm{V_{AD}}'=980$ mV.
\label{Fig2}}
\end{figure}

Figure~\ref{Fig2}(a) shows how the QD occupancy can be tuned down to the single electron regime in device A2 (similar to device A1 except for scale, see Supplementary Material). The single electron occupation was confirmed with spin filling from magnetospectroscopy, and yields an 8 meV charging energy for the last electron. The effect of $\rm{V_{AR}}$ on the tunnel rate is qualitatively visible from the charge transitions, which go from a ``smooth'' appearance at high $\rm{V_{AR}}$, when $\Gamma$ is high compared to the measurement rate, to a speckled appearance at low $\rm{V_{AR}}$, when $\Gamma$ is of the order of the measurement rate or lower \cite{Thalakulam2010}. 

We observe a gradual decrease of the AR gate capacitance to the dot, $\rm{C_{AR\text{-}dot}}$, as the reservoir fills up with electrons, as shown in the inset of Fig.~\ref{Fig2}(a) (assuming $\rm{C_{AD\text{-}dot}}$, the capacitance of the AD gate to the dot, stays constant) . The capacitance ratio $\rm{C_{AR\text{-}dot}/C_{AD\text{-}dot}}$$=-1/m$ is extracted from the slope $m$ of the transition $N=2\rightarrow3$ in the stability diagram \cite{Grabert1993}. A similar dependence of the capacitance ratio is also observed in numerical simulations, but the agreement is only qualitative, due in part to the limitations of the semi-classical simulation. We attribute this visible curvature in the dot transitions to a screening effect of the reservoir gate potential, induced by the accumulated charges in the reservoir. This specific effect therefore seems to be caused by the use of an enhancement gate connected to a ohmic contact as a tuning knob.

Device A1 also exhibits a comparable behavior as a function of the AR and AD gate voltages (see Supplementary Material). We measured the dot-reservoir tunnel rate as a function of the voltage on gate AR for device A1, along the $N=0\rightarrow1$ charge transition, as $\rm{V_{AR}}$ was compensated with $\rm{V_{AD}}$ to preserve the charge state, as shown in Fig.~\ref{Fig2}(b). Two data sets (diamond and filled circles) were taken at different voltages on a surrounding gate, $\rm{V_{{AD}^\prime}}$.  The 467 mV difference results in a 1.5 decades global offset in tunnel rates. We subtract this offset (hollow circles) to extract a single exponential dependence of $\Gamma$ with $\rm{V_{AR}}$ \cite{Borselli2015,Maclean2007}. 

From the slope of the exponential fit, we extract a gate response of $\Delta\Gamma = 5.9\pm 0.7$ decades/$\rm{V_{AR}}$, defined as the variation in dot-reservoir tunnel rate induced by a change of 1 V on gate AR, when compensated by gate AD to keep the dot chemical potential fixed. More useful for comparison between devices is when we remove the device geometry specific capacitance by converting to change in chemical potential, $\Delta\mu_{dot}$. We define the following metric:
\begin{equation}
\beta_{AR,AD}=\Delta \Gamma_{AR,AD}/\Delta \mu_{dot},
\end{equation}
where $\Delta \Gamma_{AR,AD}$ is the change in tunnel rate induced by the change in voltage on AR (and compensated by AD), $\Delta \mu_{dot}$ is the change in chemical potential caused by gate AR (equal to the chemical potential compensated by gate AD), and we call $\beta_{AR,AD}$ the \textit{tuning orthogonality}. For device A1, the above analysis leads to $\beta_{AR,AD}=0.9 \pm 0.3$ decade/meV, using the gate lever arm $\alpha_{AR}\sim0.007$ meV/mV (from $\alpha_{AD}\sim0.22$ meV/mV). We note that the chemical potential of the QD does not actually shift for a given tunnel rate  variation here, since there is a second gate compensating the chemical potential shift from the first. Therefore, care must be taken in interpreting this ratio: it does not represent the effect of a single gate on the tunnel rate, but rather the interplay of two gates acting in opposite direction on the two quantities, with unequal contributions. 

Taken individually, more positive voltages on gates AD and AR would both tend to decrease the barrier height and width, as one would expect and as shown in the conduction band edge simulations of Fig.\ref{Fig1}(d). But if one wants to keep the dot occupation fixed, and shift from high to low tunnel rates, gates AD and AR have to be swept in opposite directions. Our measurements indicate that in this case the lever of gate AR on the tunnel barrier still overcomes the opposite effect of gate AD. Furthermore, we speculate that the screening effect from charges under AR contributes to this efficiency, as it reduces the lever of gate AR on the dot occupation, but on the tunnel barrier, such that less compensation on AD is necessary to maintain charge occupation than if no screening effect was present.

The quantity $\beta_{1,2}$ can be estimated for other designs in the literature, for any pair of gates 1 and 2 used to tune the tunnel rate and compensate for changes in the dot occupation, respectively. For comparison, we estimate $\beta_{BG,AD}=1.4 \pm 0.5$ decades/meV for the case of a  dedicated barrier gate BG compensated by the dot accumulation gate AD equivalent in a Si/SiGe device \cite{Zajac2015}. This indicates a tuning orthogonality that can reach the same order of magnitude as dedicated barrier gate devices in multi-stack architectures. The single-layer split enhancement gate layout could therefore provide a wide operation range \cite{Ciorga2000} for single-electron QD devices. Details on the calculations as well as assumptions leading to the metric $\beta$ and its limitations are provided in the Supplementary material.

 \begin{figure}
 \includegraphics*[width = 1.0\columnwidth]{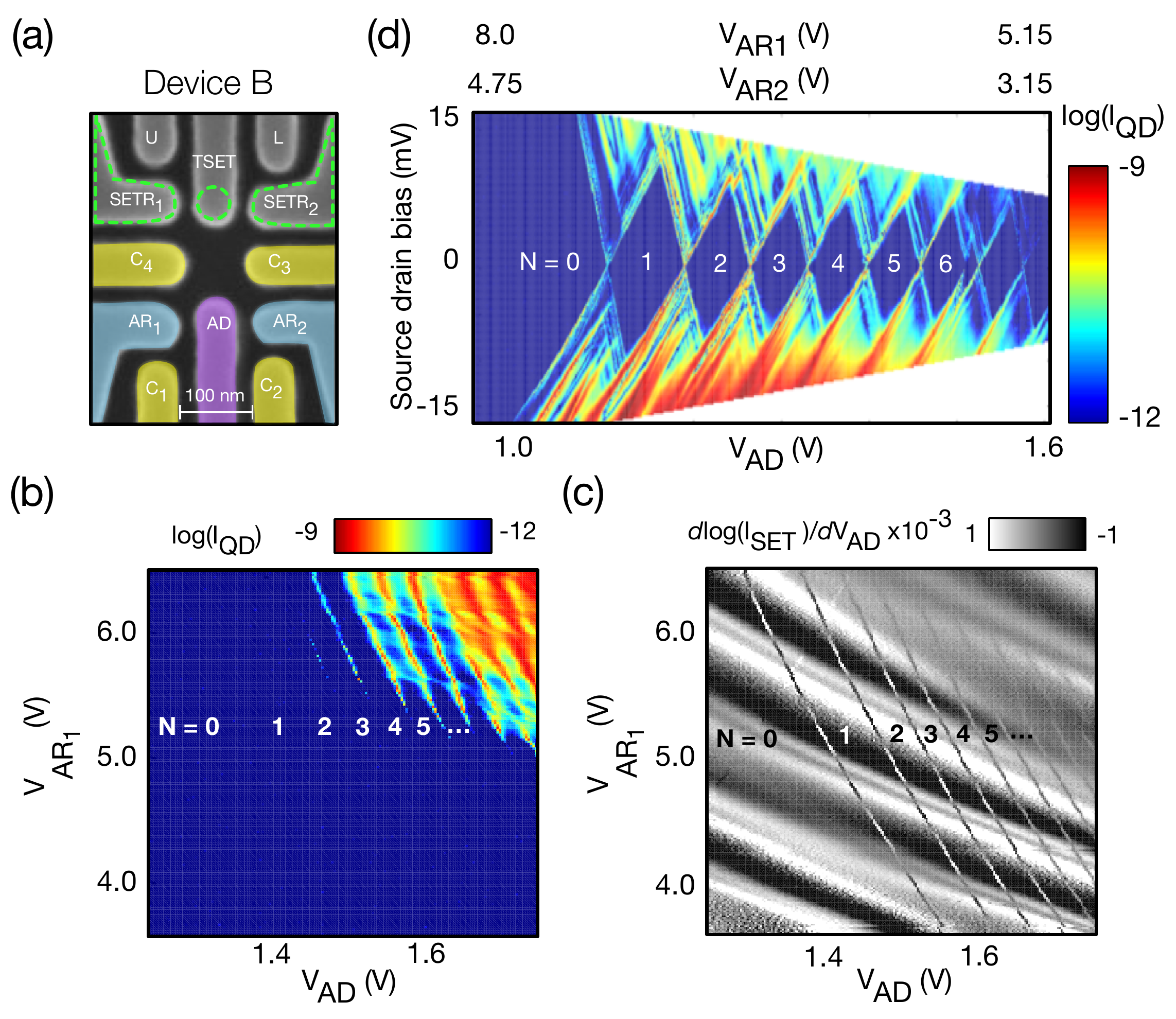}
 \caption{ (a)  SEM of a two-lead single quantum dot device, device B. C1, C2, C3 and C4 are confinement gates, AD is the dot accumulation gate, and $\rm{AR_1}$ and $\rm{AR_2}$ are the source and drain reservoir accumulation gates, respectively. A mirror structure above is operated as a SET for charge sensing.(b) Stability diagram in transport of AD vs $\rm{AR_1}$. (c) Stability diagram in charge sensing corresponding to the transport diagram in (b). (d) Coulomb diamond measurement corresponding to a stability diagram of AD vs $\rm{AR_1}$ and $\rm{AR_2}$. The small diamond after electron \#6 is due to a donor ionization \cite{Rudolph2016} (see fabrication details in the Supplementary Material).
 \label{Fig3}}
\end{figure}

The double-lead layout also supports transport down to the last electron and exhibits a typical split enhancement gate behavior. Figure~\ref{Fig3}(a) shows device B, where transport is through a QD under gate AD with source and drain reservoirs under gates $\rm{AR_1}$ and $\rm{AR_2}$.  A mirrored structure can be operated as a SET charge sensor, correlating the transport transitions (Fig.~\ref{Fig3}(b)) with charge sensed measurements (Fig.~\ref{Fig3}(c)).

In Fig.~\ref{Fig3}(b), the tunnel rate ranges from the life-time broadened regime at high $\rm{V_{AR}}$, corresponding to a $\sim$3 GHz tunnel rate \cite{Zajac2015,DeFranceschi2001} to slower than can be detected by the charge sensor, $\sim$8 Hz. The slight curvature in the dot and SET transitions of Fig.~\ref{Fig3}(d) is ascribed to a similar screening effect as in the single lead devices, although it is not as pronounced. This demonstrates that two neighboring barriers in series can be tuned relatively orthogonally (i.e., crosstalk is not a prohibitive issue), and that the split enhancement gate concept can be applied in several layouts.

In Fig.\ref{Fig3}(d), $\rm{V_{AR_1}}$ and $\rm{V_{AR_2}}$ are adjusted simultaneously to symmetrize the tunnel barriers on the source and drain side of the QD, giving rise to Coulomb diamonds \cite{Kouwenhoven1997}. The notable difference in voltage ranges applied on $\rm{AR_1}$ and $\rm{AR_2}$ is attributed mainly to asymmetry in the voltages applied on the neighboring gates on the left and right side of the device, although small variations in width of the dot-reservoir gap could also contribute to the difference. The precise effect of the dot-reservoir gap width on the tuning orthogonality and general efficiency remains to be studied in detail.
 
The addition energy of the last electron and the first orbital energy are extracted from the Coulomb diamonds of Fig.~\ref{Fig3}(d), yielding approximately $E_{add}=11$ meV and $\Delta E=3$ meV, respectively. A classical capacitance between the QD and the AD gate of 2.9 aF is estimated (e.g., $C_{AD}=e/\Delta\rm{V_{AD}}$ with $\Delta\rm{V_{AD}}=56$ meV the voltage applied on gate AD to go from the $N=0\rightarrow1$ charge transition to the $N=1\rightarrow2$ transition in Fig.~\ref{Fig3}(b)). The classical capacitance can be associated with a circular 2D QD below the gate and is used to estimate a QD radius of $\sim$30 nm, using $\epsilon_r= 3.9$ for the $\rm{SiO_2}$ and neglecting small errors due to the electron offset from the $\rm{SiO_2}$ interface and depletion of the polysilicon. The orbital energy also provides an estimate of QD size.  Following Ref. \cite{Zajac2015}, we can extract an effective length of a confining 2D box ($\pi r^2 = L^2$) and using $\Delta E=\frac{3\hbar^2\pi^2}{2m^*L^2}=$3 meV, we obtain a similar dot size, $r\sim$25 nm, using $m^* = 0.19\ m_e$. These estimated dot sizes and energies are similar to the ones obtained in multi-stack accumulation mode quantum dot devices \cite{Zajac2015,Angus2007}. 

Finally, an investigation of the spin filling and singlet-triplet energy splitting in our silicon QDs using magnetospectroscopy \cite{Borselli2011a, Lim2011a, Zajac2015} indicates that the valley splitting is linearly tunable with the vertical electric field ($8.1\pm 0.6\ \mu$eVm/MV in the double-lead device) and is tunable over a range of $\sim 75\text{-}250\ \mu$eV (see the Supplementary Material for details).


In conclusion, we explored a split enhancement gate architecture implemented in single-lead and double-lead layouts of polysilicon MOS QD devices. The single-electron regime was reliably achieved in three different devices. Using the reservoir enhancement gate to modulate the tunnel rate and compensating with the dot enhancement gate, we found a tuning orthogonality $\beta_{AR,AD}\approx 0.9$ decade/meV in one of the single-lead devices. We argue that the notable tuning orthogonality, which is comparable to what can be achieved in devices with a dedicated barrier gate in multi-stack architectures, is boosted by the screening effect arising from the use of an enhancement gate as a tuning knob. In addition, a strongly confined quantum dot with charging energies up to 11 meV and orbital energy of 3 meV was observed in the device with smallest features, corresponding to a $\sim 30$ nm radius QD. Linear tunability of the QD's valley splitting was also observed up to 250 $\mu$eV.

\section*{Supplementary Material}
Section I of the Supplementary Material provides details on the samples fabrication. Section II describes experimental details and devices characteristics. Section III presents a discussion on the tuning orthogonality metric, and Section IV is dedicated to the study of the valley splitting tuning in the split enhancement gate devices.

\begin{acknowledgments}
We gratefully recognize conversations with J. K. Gamble about early split gate designs, and J. Dominguez for supporting preparation of the devices. We acknowledge technical support from M. Lacerte, R. Labrecque, and M. Lapointe-Major, and helpful discussions with J. Camirand Lemyre, L. Schreiber and J. Klos. This work was supported by the Natural Sciences and Engineering Research Council of Canada (NSERC) and the Canada Foundation for Innovation (CFI). This research was undertaken thanks in part to funding from the Canada First Research Excellence Fund. This work was performed, in part, at the Center for Integrated Nanotechnologies, an Office of Science User Facility operated for the U.S. Department of Energy (DOE) Office of Science. Sandia National Laboratories is a multimission laboratory managed and operated by National Technology and Engineering Solutions of Sandia, LLC, a wholly owned subsidiary of Honeywell International, Inc., for the U.S. Department of Energy's National Nuclear Security Administration under contract DE-NA0003525. This paper describes objective technical results and analysis. Any subjective views or opinions that might be expressed in the paper do not necessarily represent the views of the U.S. Department of Energy or the United States Government.
\end{acknowledgments}


%

\end{document}



\title{Supplementary Material for "Quantum dots with split enhancement gate tunnel barrier control"}

\author{S. Rochette}
\thanks{S. Rochette and M. Rudolph contributed equally to this work.}
\affiliation{Institut Quantique, Universit\'e de Sherbrooke, Sherbrooke J1K 2R1, Canada}
\affiliation{D\'epartement de physique, Universit\'e de Sherbrooke, Sherbrooke J1K 2R1, Canada}

\author{M. Rudolph}
\thanks{S. Rochette and M. Rudolph contributed equally to this work.}
\affiliation{Sandia National Laboratories, Albuquerque, New-Mexico 87185, USA}

\author{A.-M. Roy}
\affiliation{D\'epartement de physique, Universit\'e de Sherbrooke, Sherbrooke J1K 2R1, Canada}

\author{M. J. Curry}
\affiliation{Department of Physics and Astronomy, University of New-Mexico, Albuquerque, New-Mexico 87131, USA}

\author{G. A. Ten Eyck}
\affiliation{Sandia National Laboratories, Albuquerque, New-Mexico 87185, USA}

\author{R. P. Manginell}
\affiliation{Sandia National Laboratories, Albuquerque, New-Mexico 87185, USA}

\author{J. R. Wendt}
\affiliation{Sandia National Laboratories, Albuquerque, New-Mexico 87185, USA}

\author{T. Pluym}
\affiliation{Sandia National Laboratories, Albuquerque, New-Mexico 87185, USA}

\author{S. M. Carr}
\affiliation{Sandia National Laboratories, Albuquerque, New-Mexico 87185, USA}

\author{D. R. Ward}
\affiliation{Sandia National Laboratories, Albuquerque, New-Mexico 87185, USA}

\author{M. P. Lilly}
\affiliation{Sandia National Laboratories, Albuquerque, New-Mexico 87185, USA}
\affiliation{Center for Integrated Nanotechnology, Sandia National Laboratories, Albuquerque, New-Mexico 87185, USA}

\author{M. S. Carroll}
\affiliation{Sandia National Laboratories, Albuquerque, New-Mexico 87185, USA}

\author{M. Pioro-Ladri\`ere}
\affiliation{Institut Quantique, Universit\'e de Sherbrooke, Sherbrooke J1K 2R1, Canada}
\affiliation{D\'epartement de physique, Universit\'e de Sherbrooke, Sherbrooke J1K 2R1, Canada}
\affiliation{Quantum Information Science Program, Canadian Institute for Advanced Research, Toronto M5G 1Z8, Canada}

\pacs{}
\maketitle

\section{Samples fabrication}
The fabrication is composed of two phases.  The first phase is run in a 0.35 micron CMOS silicon foundry, and the second phase is performed in another fabrication area that provides more flexibility in processing, particularly the e-beam lithography used for the nanofabrication.  Three different devices are presented in this work.  We describe the process flow for devices A1 and A2.  Significant differences in the structure are noted for device B.

\textit{Phase 1} (silicon foundry): The initial material stack is fabricated using a 0.35 micron silicon foundry process at Sandia National Laboratories. The starting material is a 150 mm diameter float zone <100> n-type silicon wafer with a room temperature resistivity of 10 000 $\Omega$-cm.  Device B used a p-type float zone substrate with a 99.95\% $\rm{Si^{28}}$ enriched epitaxy layer instead.  A 35 nm thermal silicon oxide is grown at 900$\degree$C with dichloroethene (DCE) followed by a 30 min, 900$\degree$C $\rm{N_2}$ anneal. The next layer deposited is a 100 nm amorphous silicon layer followed by a $5\times10^{15} \rm{cm^{-2}}$, 10 keV arsenic implant at 0$\degree$ tilt.  Device B used a 200 nm layer and the implant energy was 35 keV with the same dose.  The amorphous layers are crystallized later in the process flow to form a degenerately doped poly-silicon electrode. In the silicon foundry, the poly-Si is patterned and etched into large scale region, a ``construction zone'' around $100\ \mu\rm{m} \times100\ \mu\rm{m}$ in size, that will later be patterned using e-beam lithography to form the nanostructure.  

After etching, Ohmic implants are formed using optical lithography and implantation of As at $3\times10^{15} \rm{cm^{-2}}$ density at 100 keV.  An oxidation anneal of 900$\degree$C for 13 min and an $\rm{N_2}$ soak at 900$\degree$C for 30 min follows the implant step and serves the multiple purposes of crystallizing, activating and uniformly diffusing the dopants in the poly-Si while also forming a $\rm{SiO_2}$ layer (10-25 nm) on the surface of the poly-Si. This $\rm{SiO_2}$ layer forms the first part of the hard mask layer used for the nanostructure etch in the construction zone.  The second part of the hard mask is a 20 nm $\rm{Si_3N_4}$ layer (35 nm for device B). An 800 nm thick field oxide is subsequently deposited using low pressure chemical vapor deposition (CVD), tetraethoxysilane (TEOS) or high density plasma CVD for device B.  The field oxide is planarized using chemical mechanical polishing (CMP) leaving approximately 500 nm over the silicon and 300 nm over the poly-Si. Vias are etched to the conducting poly-Si and $\rm{n^+}$ Ohmics at the silicon surface.  The vias are filled with Ti/TiN/W/TiN. The tungsten is a high contrast alignment marker for subsequent e-beam lithography steps.  Large, approximately $100\ \mu\rm{m} \times100\ \mu\rm{m}$ windows aligned to the construction zones are then etched in the field oxide to expose the underlying hardmask and poly-Si construction zone for nanostructure patterning. The last processing step for the devices in the silicon foundry is a 450$\degree$C forming gas anneal for 90 min.

\textit{Phase 2} (separate nano-micro fabrication facility): The wafers are removed from the silicon foundry and subsequently diced into smaller parts, leading to 10 mm $\times$ 11 mm dies, each containing 4 complete QD devices. The nanostructure are patterned using electron beam lithography and a thinned ZEP resist. The pattern is transferred with a two-step etch process. First, the SiN and $\rm{SiO_2}$ hard mask layers are etched with a $\rm{CF_4}$ dry etch, followed by an $\rm{O_2}$ clean which strips the resist \textit{in-situ}. The second etch step is to form the poly-Si electrodes, which is done with an HBr dry etch in the same chamber. The poly-Si etch is monitored using end-point detection in a large scale etch feature away from the active regions of the device.  Wet acetone and dry $\rm{O_2}$ cleans are used to strip the residual resist after the poly-silicon nanostructure formation.  After the wet strips of the tungsten vias, a lift-off process is used for formation of aluminum bond pads to contact the Ohmics and poly-silicon electrodes.  

The last step is a 400$\degree$C, 30 minute forming gas anneal. For device B, after the polysilicon etch, a second e-beam lithography and implant step was done to place donors near the QD region.  The device was sent out for implant, $4\times10^{11} \rm{cm^{-2}}$ Phosphorus at 45 keV.  After the implant step, the photoresist was stripped with acetone and then the metal and residual organics were stripped from the surface using peroxide and RCA cleans.  The device was subsequently metallized using an Al lift-off process similar to devices A1 and A2.

\section{Devices and experimental parameters}

This appendix provides supplementary information on the devices and experimental parameters presented in the main text.

Experiments were performed in two distinct laboratories, Universit\'e de Sherbrooke (devices A1 and A2) and Sandia National Laboratories (device B), in dilution refrigerators sustaining electronic temperatures of 125 mK and 160 mK, respectively. In the limited testing of standard measurements, the samples were found to be robust to thermal cycles (i.e., little threshold shift) and no devices were visually altered by the long-distance shipping (e.g., damage from electrostatic discharge was not observed). The devices were also electrically stable, with the drift of the quantum dot chemical potential in device B characterized as approximately 5.3 $\pm$ 0.5 $\mu$eV standard deviation over a 150 hour period.

Table~\ref{Table1} compares the characteristics of devices A1, A2, and B. Table~\ref{Table2} exposes the experimental parameters for all measurements shown or mentionned in the main text for devices A1 and A2 (single-lead devices), while Table~\ref{Table3} does the same for device B.

\renewcommand{\arraystretch}{1.3}
 \begin{table*}
 \caption{Characteristics of measured devices. Devices A1 and A2 present the same layout, differing only in the spacing between the gates and the width of the gates (A2 gates are more closely packed than A1 gates). For comparison, we label the devices by the distance between gates AD and C2, and the distance between AD and AR tips (see Fig 1(c) of the main text). \label{Table1}}
 \begin{tabular}{|L{0.3\textwidth}|C{0.2\textwidth}|C{0.2\textwidth}|C{0.2\textwidth}|}
\hline
\textbf{Device}&\textbf{A1}&\textbf{A2}&\textbf{B}\\
 \hline
Reservoirs&Single lead&Single lead&Double lead\\ \hline
Device dimensions&AD-C2: 60 nm, \mbox{AD-AR: 100 nm}, \mbox{AD width: 100 nm}&AD-C2: 25 nm, \mbox{AD-AR: 30 nm}, \mbox{AD width: 75 nm}&AD-C2: 30 nm, \mbox{AD-AR: 20 nm}, \mbox{AD width: 50 nm}\\ \hline
Mobility &4560 $\rm{cm^2/V/s}$&4560 $\rm{cm^2/V/s}$&11600 $\rm{cm^2/V/s}$\\ \hline
Interface roughness&2.4 \AA&2.4 \AA&1.8 \AA\\ \hline
Percolation density &$6.0 \times 10^{11}$$\rm{cm^{-2}}$&$6.0 \times 10^{11}$$\rm{cm^{-2}}$&$1.6 \times 10^{11}$$\rm{cm^{-2}}$\\ \hline
Scattering charge density &$7.6 \times 10^{10}$$\rm{cm^{-2}}$&$7.6 \times 10^{10}$$\rm{cm^{-2}}$&$5.2 \times 10^{10}$$\rm{cm^{-2}}$\\ \hline
Interface correlation length&26 \AA&26 \AA&22 \AA\\ \hline
Wafer type&10 000 $\Omega$-cm, n&10 000 $\Omega$-cm, n&10 000 $\Omega$-cm, p*\\ \hline
Polysilicon gate stack thickness&100 nm&100 nm&200 nm\\ \hline
Silicon gate oxide thickness&35 nm&35 nm&35 nm\\ \hline
 \end{tabular}
 
 *Device B contains a 99.95\% Si$^{28}$ enriched epitaxy layer.
 \end{table*}

 
  \begin{table*}
 \caption{Experimental parameters for various data sets of the main text, for devices A1 and A2. \label{Table2}}
 \begin{tabular}{|C{0.1\textwidth}|C{0.15\textwidth}|C{0.17\textwidth}|C{0.15\textwidth}|C{0.2\textwidth}|C{0.15\textwidth}|}
\hline
\textbf{Data}&\textbf{Fig 1d}&\textbf{Fig 2a}&\textbf{Fig 2b, top inset}&\textbf{Fig 2b, bottom inset}&\textbf{Fig 4b}\\
 \hline
Device&A1&A2&A1&A1&A2\\  \hline
AD&1.75 V&1.25 to 1.65 V&0.840 to 0.870 V&0.790 to 0.820 V&1.25 to 1.40 V\\ \hline
AR&3.0 to 6.0 V&3.0 to 6.0 V&4.4 to 4.9 V&4.2 to 4.5 V&6.5 V\\ \hline
C1&-1.0 V&-3.0 V&-1.0 V&-1.0 V&-1.0 V\\ \hline
C2&-3.0 V&-1.4 V&-3.0 V&-3.0 V&-3.0 V\\ \hline
C3&-1.0 V&-1.4 V&-1.0 V&-1.0 V&-1.0 V\\ \hline
C4&-1.0 V&-1.0 V&-1.0 V&-1.0 V&-1.0 V\\ \hline
TSET&2.59 V&2.0 V&2.45V&2.59 V&2.0 V\\ \hline
U&-1.32V &-1.4 V&-3.19 V&-2.32V&-1.4 V\\ \hline
L&-2.06 V&-1.4 V&-1.75V&-2.06V&-1.4V\\ \hline
AD'&0.980V&0V&0.5127 V&0.980 V&0V\\ \hline
AR'&7.0 V&0 V&7.0 V&7.0 V&0V\\ \hline
C2'&-3.0 V&0 V&-1.0 V&-1.0 V&0 V\\ \hline
C3'&-1.0 V&0 V&-1.0 V&-1.0 V&0 V\\ \hline
TSET'&0 V&0 V&0 V&0 V&0 V\\ \hline
U'&0V&0 V&0 V&0 V&0 V\\ \hline
L'&0 V&0 V&0 V&0 V&0 V\\ \hline
Details &Thomas-Fermi numerical simulations.
&Charge sensing, \mbox{$f_{\rm{LI}}=16.4$ Hz} (lock-in frequency), \mbox{$\rm{V_{SD}}=100\ \mu$V} (source-drain voltage).
&Pulse spectroscopy, measured by charge sensing, \mbox{$f_{\rm{LI}}=19$ Hz}, \mbox{$\rm{V_{SD}}=100\ \mu$V}. 
&Single-shot measured by RF reflectometry, carrier wave \mbox{$f=180$ MHz}, bandwidth of \mbox{326 kHz}.
&Charge sensing, \mbox{$f_{\rm{LI}}=16.4$ Hz}, \mbox{$\rm{V_{SD}}=100\ \mu$V}.\\ \hline
 \end{tabular}
 \end{table*}

\begin{table*}
 \caption{Experimental parameters for various data sets of the main text, for device B.  All measurements are made with a Lock-In frequency of 492.6 Hz and a source drain bias of 50 $\mu$ V rms.\label{Table3}}
 \begin{tabular}{|C{0.1\textwidth}|C{0.30\textwidth}|C{0.15\textwidth}|C{0.15\textwidth}|C{0.15\textwidth}|}
\hline
\textbf{Data}&\textbf{Fig 3b and 3c}&\textbf{Fig 3d}&\textbf{Fig 4a}&\textbf{Fig 4b}\\
 \hline
Device&B&B&B&B\\  \hline
AD&1.2 to 1.8 V&0.9 to 1.6 V&1.8 V&1.21 to 1.8 V\\ \hline
AR1&3.0 to 7.0 V&5.15 to 8.0 V&5.0 V&5.0 V\\ \hline
AR2&3.5 V&3.15 to 4.75 V&3.0 V&3 to 3.1 V\\ \hline
C1&-2.7 V&-1.5 V&-6.7 to -5.3 V&-6.7 to -0.76 V\\ \hline
C2&-4.0 V&-3.0 V&-3.0 V&-3.0 V\\ \hline
C3&-0.26 V&0 V&-0.26 V&-0.26 V\\ \hline
C4&-4.2 V&-4.2 V&-4.2 V&-4.2 V\\ \hline
TSET&2.61 V&0 V&2.53 V&2.53 V\\ \hline
SETR1&2.5 V&0 V&2.5 V&2.5 V\\ \hline
SETR2&2.5 V&0 V&2.5 V&2.5 V\\ \hline
U&-1.5 V&0 V&-4.8 V&-4.8 V\\ \hline
L&-4.8V&0V&-0.92V&-0.92 to -1.26 V\\ \hline
 \end{tabular}
 \end{table*}
 


\begin{figure*}
 \includegraphics*[width = 1.6\columnwidth]{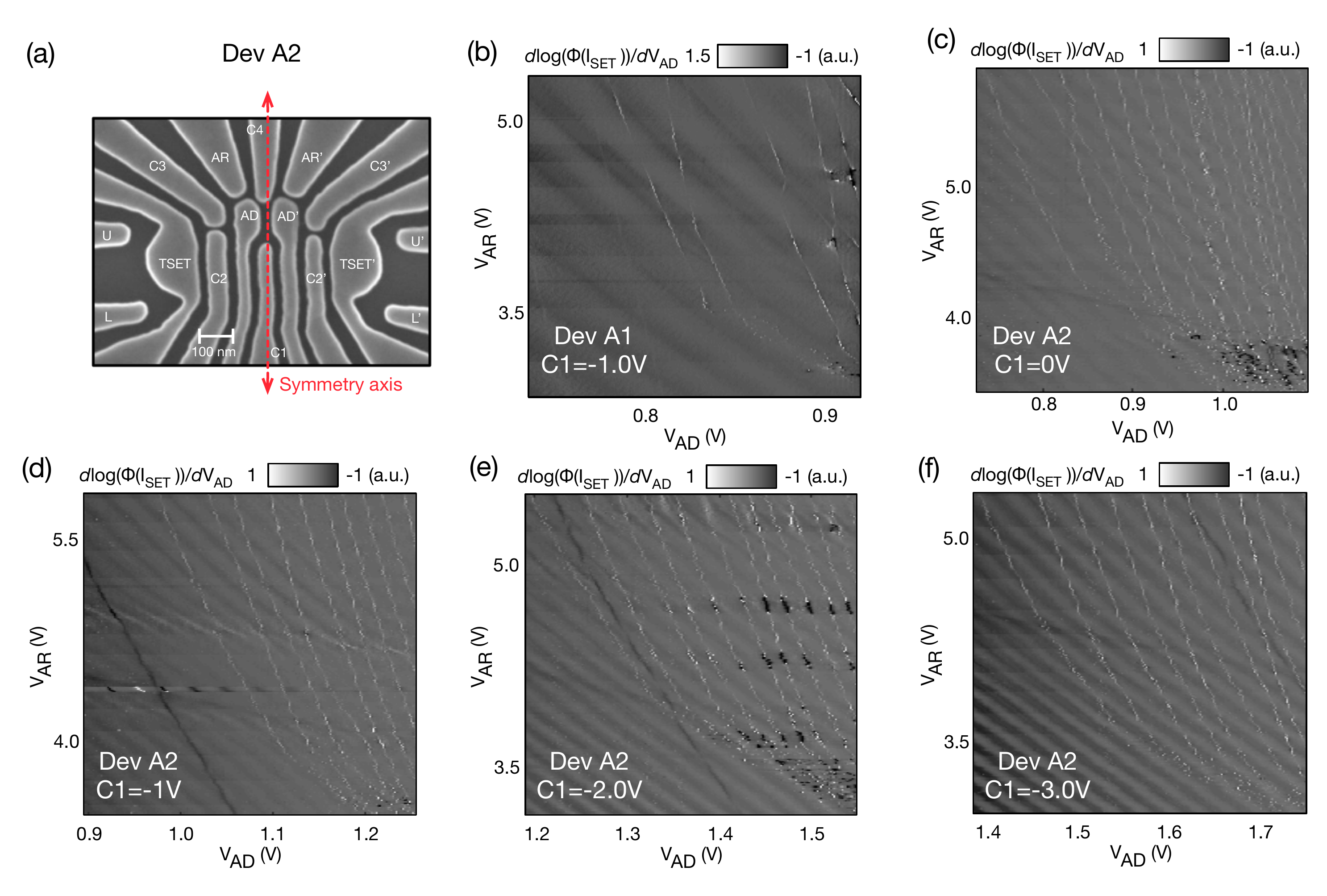}
 \caption{(a) SEM of single-lead device A2. The device has a symmetry axis between the two quantum dots. Experiments on device A2 involved the formation of a single quantum dot, on the left side of the device only (under AD).(b) Wide range stability diagram for device A1 corresponding to bottom right inset of Fig. 2(b) in the main text. The pale charge transitions on the left-hand side are transitions in the left QD, which was activated for this measurement series. The large features of device A1 and the small negative voltage on C1 are responsible for the irregularities in the right dot transitions (right hand side). (c), (d), (e), (f) Stability diagrams for device A2, with all parameters kept the same except for C1 gate voltage. A more negative voltage on C1 leads to more regular quantum dots, as expected. \label{Fig5}}
\end{figure*}

A statement concerning device A1 is helpful for full comprehension. The full range AD vs AR stability diagram for device A1 is not shown in the main text for the sake of clarity. Indeed, features not related to the split enhancement gate operation principles, and attributed to an irregularly shaped confinement potential under gate AD, were present in the full-range stability diagrams of device A1 (see Fig.~\ref{Fig5}(b)). This effect could be mitigated, but only up to a certain point, by applying more negative voltages on gates C1 and C2. The stability diagram of device A2 however is much cleaner owing to its smaller features compared as to A1, but experimental setup constraints at that time prevented us from repeating the tunnel rate measurements on device A2, hence why we rely on qualitative analysis only for this device. We emphasize that with the appropriate confinement, both devices qualitatively exhibit the same tunnel rate modulation and bending of the charge transitions, which, as stated in the main text, we believe is intrinsic to the split enhancement gate tunnel barrier.

Figure \ref{Fig5}(b) illustrates the effect of an insufficient and irregular confinement of the dot in device A1. Figures \ref{Fig5} (c)-(f) show how the smaller features of device A2, combined with an increasingly more negative voltage on gate C1, lead to more regular dot transitions and the clean diagram shown in Fig.~2(a) of the main text. This observation is in agreement with the clean and regular transitions witnessed for device B (Fig.~3(d) of the main text), which possesses even smaller features than device A2 (see Table \ref{Table1}).

\section{Tuning orthogonality}

When designing a QD device, it is of interest to provide local control of important device properties, with the surface gate voltages often serving as the control knobs.  One oft used parameter is gate lever arm $\alpha$, which describes the efficacy of a gate voltage on the QD chemical potential level $\mu$. The lever arm is defined as
\begin{equation}
\Delta\mu_i=\alpha_i\Delta V_i
\end{equation}
where there is a unique $\alpha_i$ for each gate $i$. In a similar spirit, a parameter describing the controllability of the QD-reservoir tunnel rate can be defined as
\begin{equation}
\Delta\Gamma_i=\beta_i'\Delta V_i
\end{equation}
While $\alpha$ is always positive by definition, $\beta'$  can be positive or negative, depending on if gate $i$ increases or decreases the reservoir-QD tunnel rate with a positive voltage change.  For example, for a QD under gate AD, gate AR increases the tunnel rate with increasing voltage, while gate AD$'$ decreases the tunnel rate with increasing voltage (Fig. 2(b) of the main text). Geometric arguments can typically be made to estimate the sign of $\beta'$ by considering whether a positive voltage change on a gate is pulling the dot towards or away from the reservoir.

Of particular interest for designing QDs is the ability to tune the tunnel rates to the QD while only imparting a minimal change in the QD chemical potential, which denotes a high degree of tuning orthogonality between the two properties. Good orthogonality facilitates emptying the QD (fewer gate compensations are required to obtain $N=1$) and tuning the reservoir coupling with minimal effect on the location of charge transitions in the stability diagram (quicker optimization of relaxation and coherence times).  For a single gate, the orthogonality between the tunnel rate and the chemical potential tunability is optimized by maximizing the ratio $\frac{\Delta\Gamma_i}{\Delta\mu_i}=\frac{\beta_i'}{\alpha_i}\equiv\beta_i$.  We rewrite this in an analogous form to the lever arm:
\begin{equation}
\Delta\Gamma_i=\beta_i\Delta\mu_i
\end{equation}
To obtain $\beta_i$, one must measure the change in both tunnel rate and chemical potential for a change on the gate voltage $\Delta V_i$.  In practice, this is impossible because a change in a single voltage moves the QD level out of resonance with the Fermi level, and a change in tunnel rate cannot be determined.  Thus, one must consider the effect of two gate voltages changing and compensating each other such that the QD chemical potential is always in resonance with the Fermi level.  Continuing the analogy with the lever arm, we assume that the total change in tunnel rate is simply the sum of the contributions of each gate that has changed.  For two gates 1 and 2, this results in
\begin{equation}
\Delta\Gamma_{1,2}=\Delta\Gamma_{1}+\Delta\Gamma_{2}=\beta_1\alpha_1\Delta V_1+\beta_2\alpha_2\Delta V_2 .
\label{equ4}
\end{equation}
As the chemical potential has not changed, we have the additional constraint
\begin{equation}
\Delta\mu_{1,2}=\Delta\mu_{1}+\Delta\mu_{2}=\alpha_1\Delta V_1+\alpha_2\Delta V_2=0 .
\label{equ5}
\end{equation}
Combining Eq.~\ref{equ4} and Eq.~\ref{equ5}, we can define the two-gate tunnel rate orthogonality parameter as
\begin{equation}
\beta_{1,2}\equiv\beta_{1}-\beta_{2}=\frac{\Delta\Gamma_{1,2}}{\Delta\mu_1},
\end{equation}

which is directly attainable from the measurements in Figure 2 (b) of the main text.  From the data, we extract a slope of $\frac{\Delta\Gamma_{AR,AD}}{\Delta V_{AR}}=5.9 \pm 0.7$ decades/$\rm{V_{AR}}$, describing the change in tunnel rate induced by a change in both $\rm{V_{AR}}$ and $\rm{V_{AD}}$.  With a lever arm $\alpha_{AR}\sim0.007$ eV/V, we determine $\beta_{AR,AD}=0.9 \pm 0.3$ decades/meV.


For comparison, we extract $\beta_{1,2}$ for a multilayer enhancement mode Si/SiGe device which uses a dedicated barrier gate located directly on top of the tunnel barrier, sandwiched between the reservoir and QD gates (Zajac \textit{et al.}\cite{Zajac2015}). Information on the tunnel rates is determined from the stability diagram of the tunnel barrier gate LB1 and the QD gate L1 (Fig. 2a of Zajac \textit{et al.}\cite{Zajac2015}).  To more easily compare this data to our device, we relabel LB1$\rightarrow$BG and L1$\rightarrow$AD.  The voltage ranges studied show transition rates ranging from the measurement sample rate (assumed to be at least 10 Hz) to the lifetime broadened regime ($\frac{k_BT_e}{h}=800$ MHz for a reported electron temperature of $T_e=40$ mK).  This provides two coordinates $(\Gamma,\rm{V_{BG}})$  to estimate the tunnel rate orthogonality, for which we find $\Delta\Gamma_{BG,AD}=\frac{7.9 \rm{decades}}{0.4 \rm{V_{BG}}}=19.8\ \rm{decades/V_{BG}}$.  From the reported lever arms and capacitance ratio for the QD and barrier gates, we determine $\alpha_{BG}=0.022$ eV/V, and thus $\beta_{BG,AD}=1.4 \pm 0.5$ decades/meV. 


The definition of $\beta_{1,2}$ lends itself to compare other devices and geometries as well, as $\beta_{1,2}$ is independent of geometry specific information like capacitances.  The concept of $\beta_{1,2}$ can also be extended to optimize QD devices for other characteristics which may be useful for qubit operation.  For example, one can similarly define a parameter that describes the orthogonality between a double-QD coupling and the double-QD detuning, or a double-QD coupling and the valley splitting.

\section{Valley splitting tuning}

In this section, we examine the spin filling and singlet-triplet energy splitting in our silicon QDs using magnetospectroscopy \cite{Borselli2011a, Lim2011a, Zajac2015}.  

\begin{figure}
	\includegraphics*[width = 1.00\columnwidth]{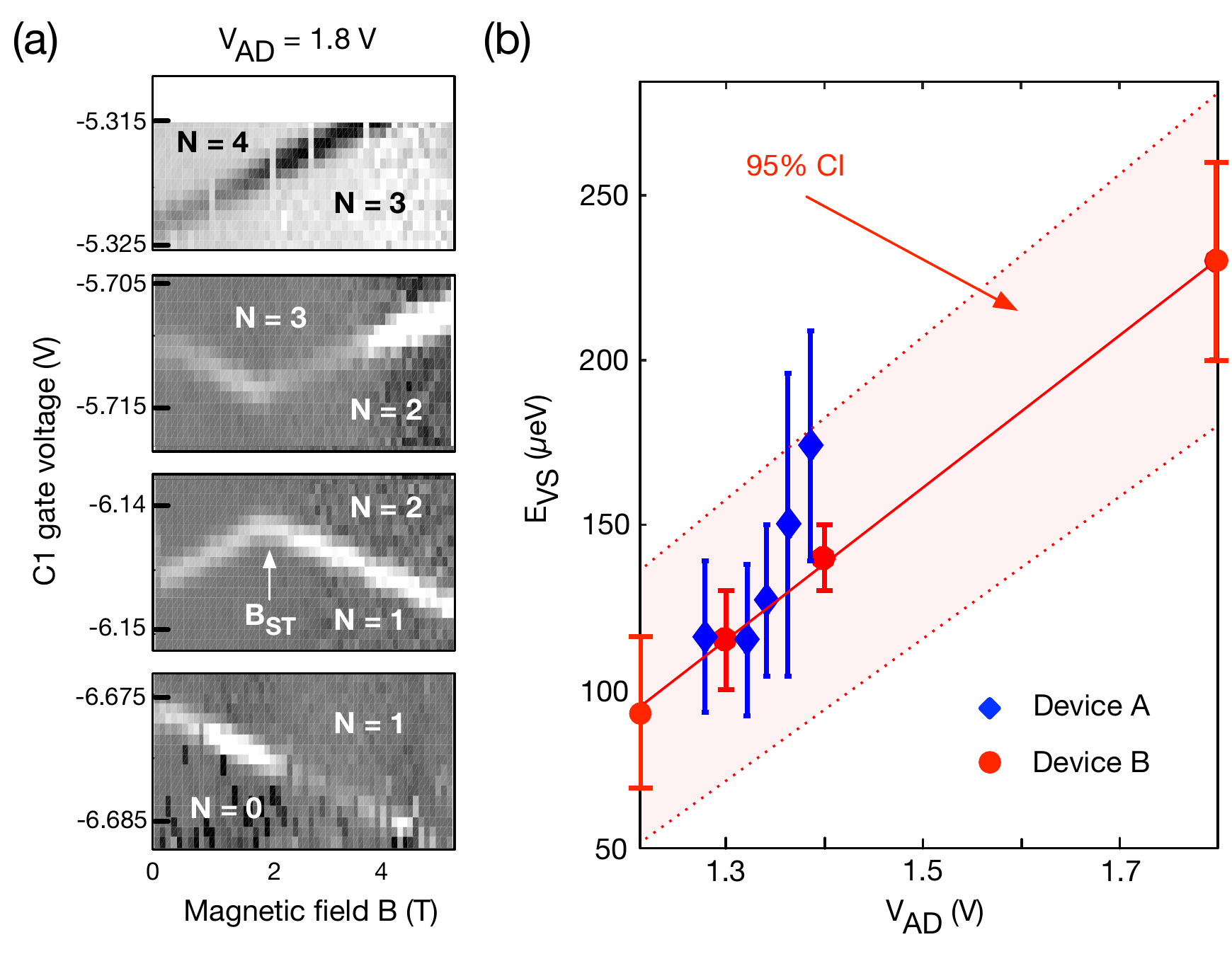}
	\caption{ (a) In-plane magnetospectroscopy measurements for device B, for transitions $N=0\rightarrow$1, $1\rightarrow2$, $2\rightarrow3$, and $3\rightarrow4$, from a stability diagram similar to Fig. 3(c) of the main text, at $\rm{V_{AD}}$=1.8 V. A lever arm of $31\pm4\ \mu$ eV/mV is inferred assuming $\mbox{g=2}$, within 15\% of the lever arm extracted from Coulomb peak width temperature dependence \cite{Beenakker1991}. $B_{ST}$ indicates the magnetic field at which the singlet-triplet transition occurs. (b) Extracted valley splitting $E_{VS}$ as a function of the dot accumulation gate voltage $\rm{V_{AD}}$. The diamond (blue) data points are for device A2 (single-lead, Fig.~\ref{Fig5}), and the circle (red) data points are for device B (two-leads, Fig. 3(a) of the main manuscript). The dashed red line indicates the fit for the valley splitting tunability of device B, and the 95\% confidence range (CI) is indicated by the red filled region.\label{Fig4}}
\end{figure}

The first 4 charge transitions from device B are shown as a function of transverse magnetic field, at $\rm{V_{AD}}$=1.8 V, in \ref{Fig4}(a).  The first transition shows a shift in chemical potential consistent with a lowering of energy due to increasing Zeeman splitting. The inflection point at $B=B_{ST}$ in the  $N=1\rightarrow 2$ charge transition indicates the magnetic field at which the singlet-triplet (ST) transition occurs in the quantum dot \cite{Yang2013,Culcer2010}. The magnetospectroscopy for the $N=2\rightarrow 3$ transition has an inflection at the same B-field as the $N=1\rightarrow 2$  transition.  This is consistent with a simple model for which there are two valleys and the $\rm{2^{nd}}$  valley is loaded with a $\rm{3^{rd}}$ electron as spin down. The inflection point again marks the crossing of the spin up of the lower valley with the spin down of the upper valley. The $\rm{4^{th}}$  electron then loads always spin up, also suggesting that the next orbital energy is well offset from this lower manifold, which is indeed consistent with the order of 3 meV estimate from the Coulomb diamonds. This spin filling also indicates a relatively small Coulomb repulsion relative to orbital energy spacing \cite{Hada2003}.

The magnetospectroscopy measurements were repeated for different $\rm{V_{AD}}$, compensating with the confinement gate C1 to maintain charge occupation. We estimated the single particle valley splitting from  $E_{VS}=g\mu_B B_{ST}$, assuming $g=2$, for devices A2 and B (Fig.~\ref{Fig4}(b)). For device B, we extract a linear tunability of $E_{VS}$ with the accumulation gate voltage of $231\pm 15\ \mu$ eV/V, the error range corresponding to a 95\% confidence interval on the fit. Roughly approximating the vertical electric field as $\Delta F_Z=\Delta \rm{V_{AD}}$/$t_{ox}$, where $t_{ox}$ is the gate oxide thickness, 35 nm here, we convert this tunability to $8.1\pm0.6\ \mu$eV m/MV. The linear trend is qualitatively consistent with theory and recent observations in MOS QDs \cite{Gamble2016,Yang2013}. 

For device A2, although the measurements were too noisy to extract a convincing tunability fit, all data points are located within the confidence interval for device B's tunability. We note that differences in valley splittings between devices A2 and B would be expected from variations in electrostatic environments (e.g., gate layout and dimensions,  distribution of voltages to reach single electron occupation and threshold voltages) and in interface roughness, approximately 20\% different between the two samples \cite{Gamble2016}. 

%

